# Supersoft X-Ray Sources in M 31

J. Greiner[1], R. Supper[1], E.A. Magnier[2,3]

[1] Max-Planck-Institute for Extraterrestrial Physics, 85740 Garching, Germany
[2] University of Amsterdam, Astronomy Dept, 1098 SJ Amsterdam, The Netherlands
[3] University of Washington, Dept of Astronomy FM-20, Seattle, WA 98195, USA

## 1 Introduction

As the most massive galaxy in the Local Group the early-type spiral M31 is close enough to be studied in detail at various wavelengths. A ROSAT PSPC mosaic of 6 contiguous pointings with an exposure time of 25 ksec each was performed in July 1991 (first M31 survey). A raster pointing of 80 observations with 2.5 ksec exposure time each covering the whole M31 disk was made in July/August 1992 and January 1993 (second M31 survey). A multi-step detection algorithm involving a maximum likelihood technique as final step was applied to the individual pointings of the first survey with a likelihood threshold of 10. A total of 396 X-ray sources were detected within the field of view of about 6.3 $deg^2$ (Supper et al. 1996). For each source several quantites are determined such as the position, the total number of background subtracted counts, the corresponding countrate and two hardness ratios for a crude spectral characterisation.

CCD photometry has been performed in four passbands (*BVRI*) of the entire optical disk of M31 in September 1990 and September 1991 using the McGraw-Hill 1.3m telescope of the Michigan-Dartmouth-MIT Observatory at Kitt Peak (Magnier et al. 1992, Haiman et al. 1993). The observations have typical completeness limits of (22.3,22.2,22.2,20.9). Different selection criteria on these data sets allow the extraction of different classes of objects. Of particular interest are likely Galactic foreground stars which have been identified with the criteria V$\leq$18 and B–V>0.4 (predominantly main sequence and giant stars).

## 2 Selection Criteria and Source Sample

Two selection criteria were applied to the total sample of detected X-ray sources:

- the hardness ratio has to satisfy HR1 + $\sigma_{\rm HR1} \leq -0.80$. The hardness ratio HR1 is defined as the normalized count difference ($N_{50-200}$ – $N_{10-40}$)/($N_{10-40}$ + $N_{50-200}$), where $N_{a-b}$ denotes the number of counts in the PSPC between channels a and b. Similarly, HR2 is defined as ($N_{91-200}$ – $N_{50-90}$)/$N_{50-200}$. We note that HR1 is sensitive to the absorbing column, and that HR2 is basically zero (within the errors) for supersoft sources. The error in the hardness ratio ($\sigma_{\rm HR1/2}$) was calculated by applying Gaussian error propagation.
- no correlation with any optical object in the error box which has been classified as a foreground object.

**Table 1.** Summary of the supersoft X-ray sources in M31 according to our selection criteria (see text for details) plus the source proposed by White et al. (1995) as supersoft transient (last row). Given for each source are the name (column 1), the best fit X-ray position (2), the $3\sigma$ location error (3), the total number of counts collected in the corresponding pointing (exception: for the White et al. source all pointings are used; 4), the PSPC countrate in the 0.1-2.4 keV band (5), the hardness ratio HR1 with error (6), the hardness ratio HR2 with error (7), and the maximum blackbody temperature (8, see text).

| Name | Coordinate (2000.0) | Error ($''$) | $N_{cts}$ | countrate (cts/ksec) | HR1 | HR2 | $T_{bb}^{max}$ (eV) |
|---|---|---|---|---|---|---|---|
| RX J0037.4+4015 | $00^h37^m25\overset{s}{.}3$ $+40°15'16''$ | 18 | 10 | 0.31±0.31 | −0.93±0.07 | 0.02±0.71 | 43 |
| RX J0038.5+4014 | $00^h38^m32\overset{s}{.}1$ $+40°14'39''$ | 33 | 33 | 0.80±0.28 | −0.92±0.08 | −0.49±0.53 | 45 |
| RX J0038.6+4020 | $00^h38^m40\overset{s}{.}9$ $+40°20'00''$ | 15 | 74 | 1.73±0.29 | −0.93±0.06 | 0.32±0.66 | 43 |
| RX J0039.6+4054 | $00^h39^m38\overset{s}{.}5$ $+40°54'09''$ | 21 | 22 | 0.44±0.44 | −0.92±0.07 | −0.04±0.71 | 45 |
| RX J0040.4+4009 | $00^h40^m26\overset{s}{.}3$ $+40°09'01''$ | 27 | 31 | 0.85±0.32 | −0.94±0.06 | −0.90±0.10 | 42 |
| RX J0040.7+4015 | $00^h40^m43\overset{s}{.}2$ $+40°15'18''$ | 21 | 50 | 1.26±0.32 | −0.94±0.06 | −0.31±0.64 | 42 |
| RX J0041.5+4040 | $00^h41^m30\overset{s}{.}2$ $+40°40'04''$ | 18 | 16 | 0.32±0.18 | −0.95±0.05 | −0.62±0.44 | 40 |
| RX J0041.8+4059 | $00^h41^m49\overset{s}{.}9$ $+40°59'21''$ | 27 | 23 | 0.49±0.24 | −0.93±0.07 | −0.63±0.43 | 43 |
| RX J0042.4+4044 | $00^h42^m27\overset{s}{.}6$ $+40°44'32''$ | 15 | 72 | 1.69±0.32 | −0.93±0.07 | −0.07±0.70 | 43 |
| RX J0043.5+4207 | $00^h43^m35\overset{s}{.}9$ $+42°07'30''$ | 15 | 57 | 2.15±0.55 | −0.92±0.08 | −0.27±0.66 | 45 |
| RX J0044.0+4118 | $00^h44^m04\overset{s}{.}8$ $+41°18'20''$ | 15 | 69 | 2.46±0.42 | −0.94±0.06 | 0.11±0.81 | 42 |
| RX J0045.5+4206 | $00^h45^m32\overset{s}{.}3$ $+42°06'59''$ | 24 | 86 | 3.14±0.34 | −0.89±0.07 | −0.29±0.65 | 42 |
| RX J0046.2+4144 | $00^h46^m15\overset{s}{.}6$ $+41°44'36''$ | 15 | 63 | 2.15±0.39 | −0.93±0.07 | 0.62±0.40 | 38 |
| RX J0046.2+4138 | $00^h46^m17\overset{s}{.}8$ $+41°38'48''$ | 27 | 30 | 1.12±0.40 | −0.91±0.09 | −0.27±0.65 | 40 |
| RX J0047.6+4205 | $00^h47^m38\overset{s}{.}5$ $+42°05'07''$ | 30 | 29 | 1.05±0.36 | −0.92±0.07 | 0.06±0.70 | 39 |
| RX J0045.4+4154 | $00^h45^m29\overset{s}{.}0$ $+41°54'08''$ | 18 | 1040 | 29.63±0.98 | +0.78±0.03 | −0.59±0.03 | 128 |

Among all 396 detected sources a total of 15 fulfill these two requirements. These sources are listed in Tab. 1 and are identical to the sources #3, #12, #18, #39, #78, #88, #114, #128, #171, #245, #268, #309, #335, #341, and #376 in Supper et al. (1996). Fig. 1 shows the positions of these sources overplotted on an optical image of M31.

## 3 X-Ray Characteristics

We have applied a spectral fit only to the brightest source of our sample, namely RX J0045.5+4206. A blackbody model fit gives an absorbing column of $1.2\times10^{21}$ cm$^{-2}$, a factor of two higher than the Galactic value ($0.6\times10^{21}$ cm$^{-2}$, Dickey and Lockman 1990). But the uncertainty is large, and the line of sight Galactic absorption is well within the $3\sigma$ contour (see Fig. 2). The resulting blackbody temperature is $30^{+20}_{-10}$ eV at the Galactic absorbing column, similar to the known supersoft sources (Greiner et al. 1991, Kahabka et al. 1994). The bolometric luminosity for this model blackbody is about $7\times10^{37}$ erg/s. A comparison of blackbody and white dwarf atmosphere models has shown that while the temperature estimates did not differ much, the bolometric luminosity can be a factor 10–100 lower when using a white dwarf atmosphere model (Heise et al. 1994). We there-fore think that the above given bolometric luminosity derived at the Galactic absorbing column should be a better representation than the best-fit value.

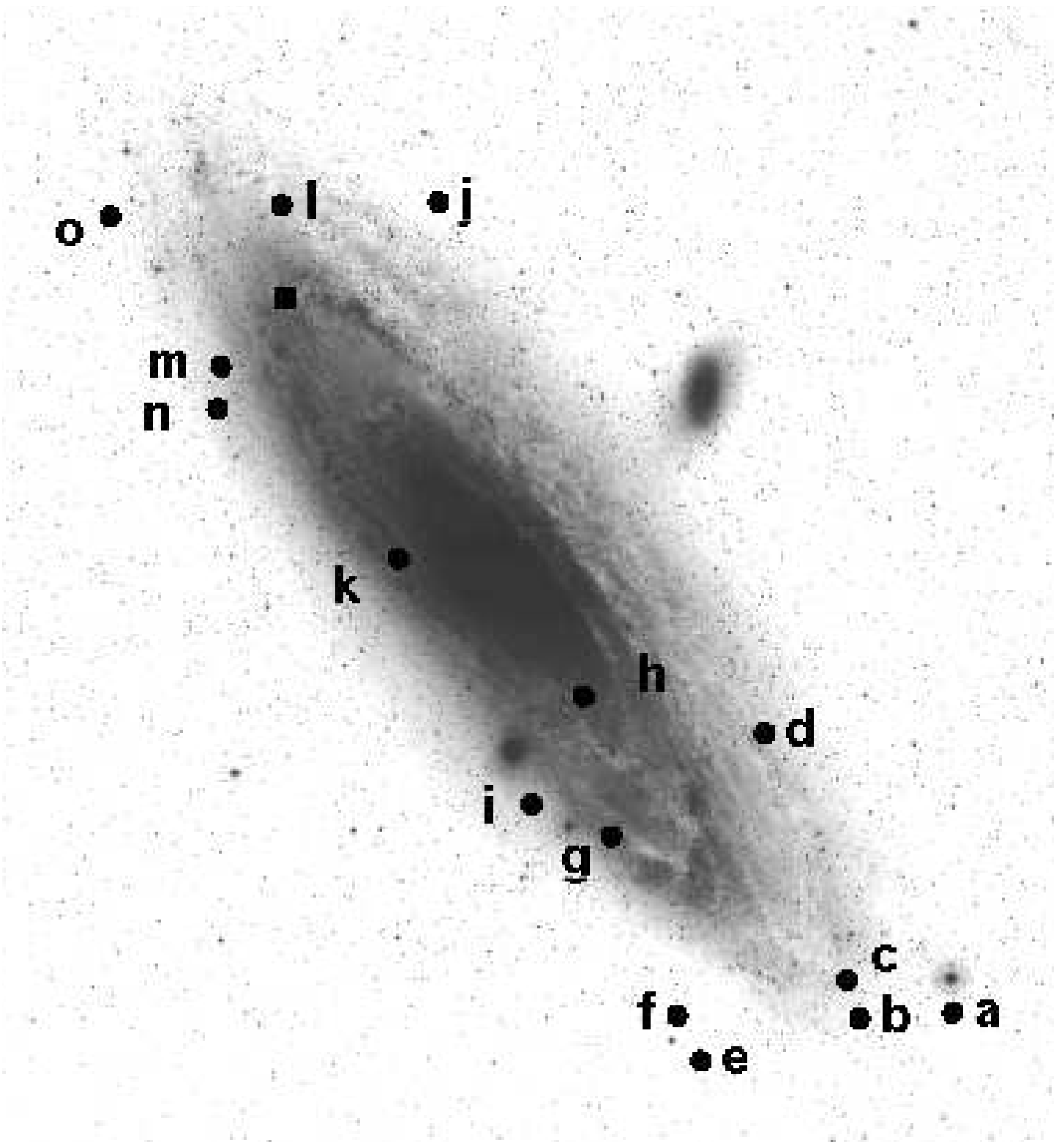

**Fig. 1.** The location of the 15 supersoft sources plotted over an optical image (POSS O plate) of M31. Letters (a through o) denote the X-ray sources in Tab. 1 from top to bottom. The source RX J0045.4+4154 (White et al. 1995) is shown as a square. Sources h, k and l have objects brighter than 20th mag in their X-ray error box (see text for possible counterpart types). Sources e and m lie at the edge of the *BVRI* survey field, while a, j and n are outside of this survey.

In order to get some crude spectral information also for the other, fainter X-ray sources we have adopted the following scheme: We calculate the hardness ratio HR1 for blackbody models with different temperatures and absorbing columns (Fig. 3). We then take the measured HR1 value, and determine the model blackbody temperature at the Galactic absorbing column. Since for a given hardness ratio the temperature decreases with increasing $N_H$, these temperatures are upper limits. As could be expected already from the rather similar values of HR1, these maximum blackbody temperatures lie in a quite narrow range between 40–50 eV (last column in Tab. 1).

Recently, White et al. (1995) have reported on a recurrent X-ray transient which they claim to be supersoft. We have determined the maximum temper-ature with the same method as our source sample, using the merged pointings of the second PSPC survey of M31. This temperature and the hardness ratio of this source (Tab. 1) clearly demonstrate that it is considerably harder than the sources we have extracted here. From the spectral similarity we speculate that it is rather a Her X-1 like object than a supersoft X-ray source.

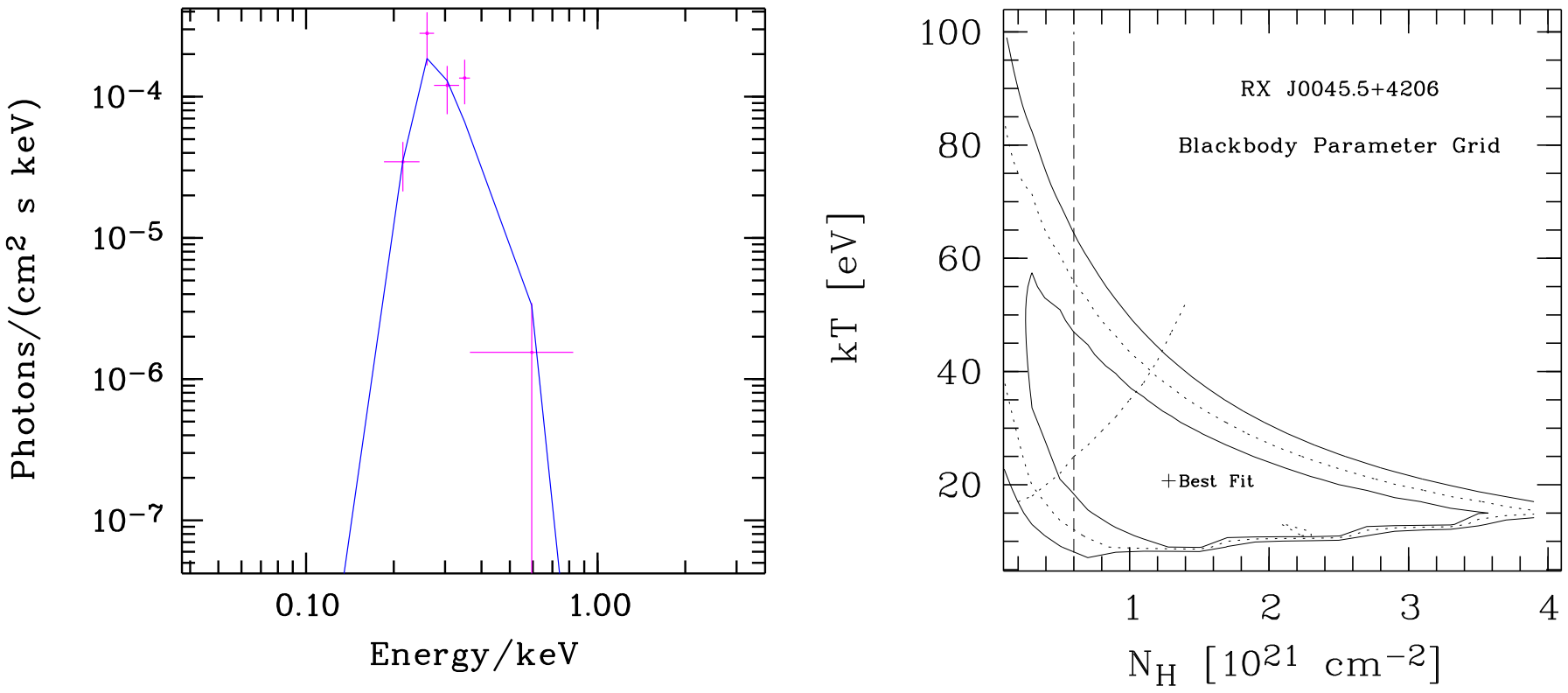

**Fig. 2.** Best fit blackbody model of the brightest of our supersoft X-ray sources RX J0045.5+4206 (left) and the $1, 2, 3\sigma$ error contours in the $kT - N_H$ plane (right panel). Only photons from one single 25 ksec pointing are used. Some photons had to be discarded because of another bright source at only 1ʹ.8 distance making the flux somewhat uncertain. Due to small photon numbers, the parameter range is not too well constrained. and thus introduces an additional uncertainty in the luminosity. In the right panel the dashed vertical line corresponds to the mean total Galactic absorption, and the dotted line marks the Eddington limit of the blackbody model.

All 15 X-ray sources have been checked for temporal variability against the August 1992 observations of the second M31 survey. None of the sources fulfilled the criterion for variability (flux difference in the two observations divided by the quadratic sum of the flux errors has to be larger than $3\sigma$).

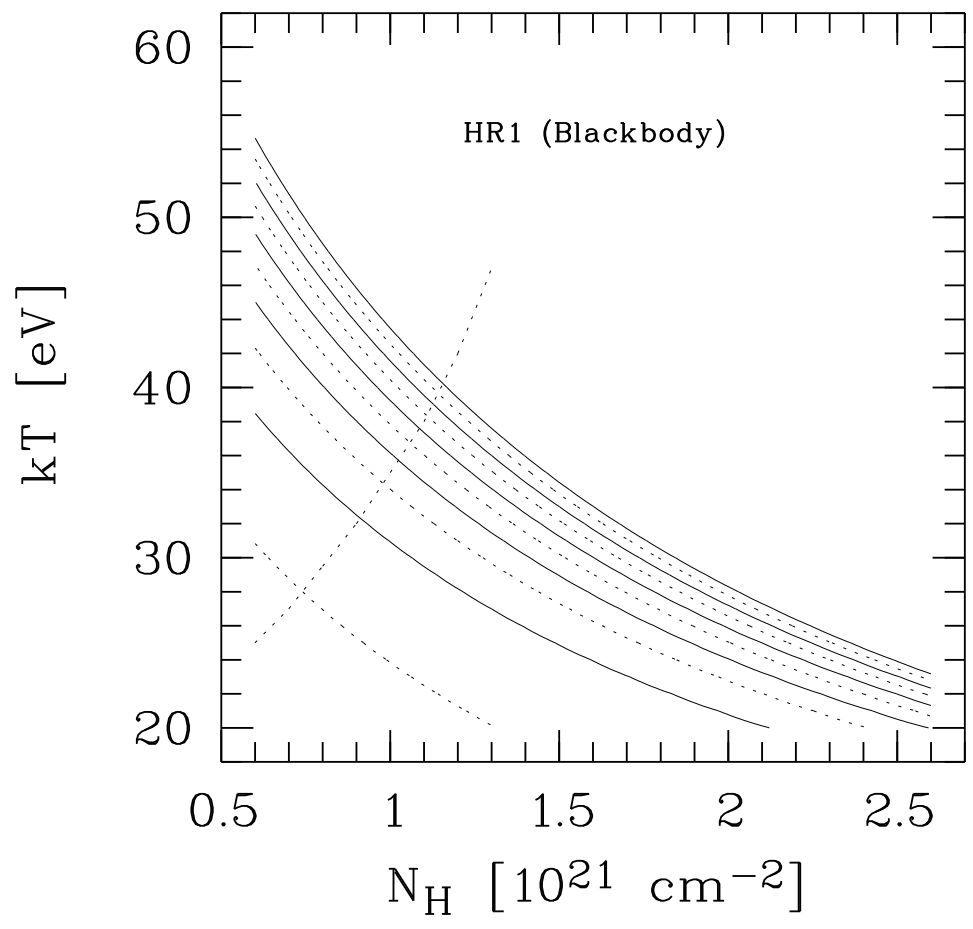

**Fig. 3.** The hardness ratio HR1 as a function of blackbody temperature kT and absorption column density $N_H$. HR1 increases from the upper line with HR1= $-0.80$ to the bottom line with HR1= $-0.98$ in steps of 0.02. For a given hardness ratio the blackbody temperature decreases with increasing column density. Therefore, the kT value at $N_H = 0.6 \times 10^{21}$ cm$^{-2}$ gives the maximum blackbody temperature. The dotted line marks the Eddington limit of the blackbody model.

## 4 Source Positions and Identifications

The 396 X-ray source positions as they result from the maximum likelihood detection algorithm have varying errors depending on the off-axis angle in the field of view of the PSPC. Since for nearly a quarter of X-ray sources optical counterparts could be identified due to the positional coincidence plus the colour information from the *BVRI* survey, we have used a subset of these optically identified X-ray sources (namely the 29 globular clusters) to improve the systematic errors in the X-ray source determination. We have determined the matrix coefficients for a unique rotation plus translation which gives the smallest residual position differences for the optical and X-ray positions of the optically identified X-ray sources. The same matrix has then been applied to all X-ray sources.

There are several source positions (Fig. 4) with optical objects inside the error box. The optical selection criterion is effective mainly against red objects of some of our supersoft sources. Single, i.e. non-interacting white dwarfs (WDs) and many magnetic cataclysmic variables (CVs) also have supersoft X-ray spectra, and such objects are blue. From the known ratio of optical to X-ray luminosity and the observed X-ray fluxes we might expect Galactic, single WDs at typical V magnitudes of 18–20 mag and CVs at even fainter magnitudes. These type of objects thus constitute a possible contamination of our sample. In particular, objects RX J0041.8+4059, RX J0044.0+4118 and RX J0045.5+4206 have blue objects of 19–20th mag in their respective error box, and thus might be Galactic WDs or CVs. On the other hand, blue stars in M31 are not distributed randomly, but in the spiral arms. The above mentioned objects are in regions with many blue stars, i.e. in associations. Thus, the probability of having a coincident CV or WD in these areas is not the sky density of CVs or WDs times the area of the survey, but rather only times the total area of the associations, which is substantially smaller. Spectroscopic observations are clearly necessary to distinguish between these alternatives.

There is recent evidence that also some active galactic nuclei (AGN) have supersoft X-ray spectra (Greiner et al. 1996). At least one AGN (WPVS007, Grupe et al. 1995) has HR1<–0.8, and there are a number of further secure identifications with HR1<–0.6. We therefore caution that one or the other source in our list which is located in the outskirts of M31 might be an AGN shining through M31. RX J0040.4+4009 and RX J0040.7+4015 are plausible candidates.

In addition to the *BVRI* survey, images were also aquired in the H$\alpha$ line. Only five of the sources turned out to lie in the H$\alpha$ fields: RX J0040.7+4015, RX J0041.5+4040, RX J0041.8+4059, RX J0042.4+4044 and RX J0044.0+4118. None of these reveals any H$\alpha$ emission. The upper limits in H$\alpha$ luminosity are $10^{35}$ erg/s, i.e. a CAL 83 like nebula (Pakull & Motch 1989, Remillard et al. 1995) would have been detected. This result is not surprising in view of the fact that in a search for nebulae around known supersoft sources in the Galaxy and the Magellanic Clouds only one out of 10 sources has revealed a nebula (Remillard et al. 1995).

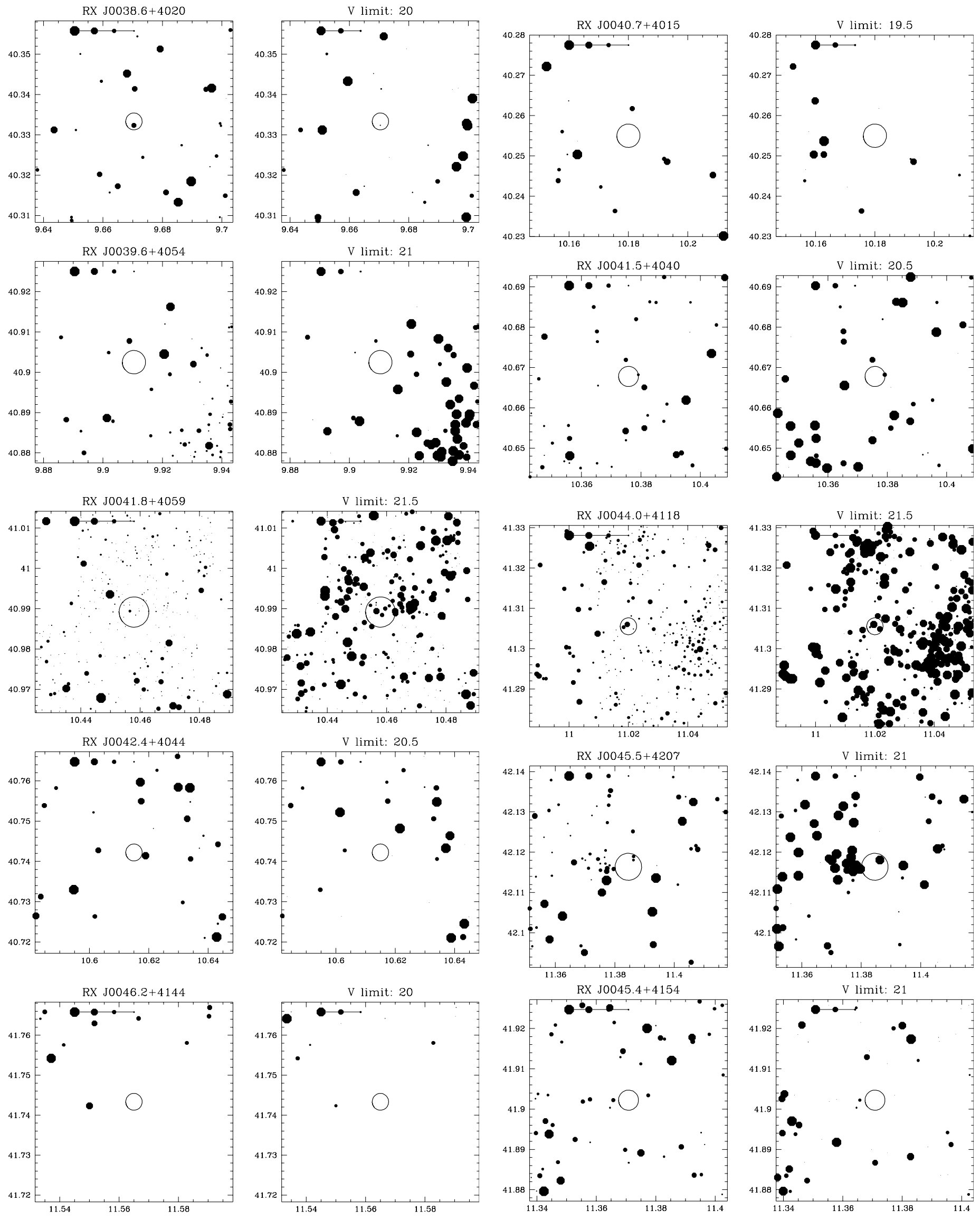

**Fig. 4.** Finding charts for those supersoft sources covered by the *BVRI* survey which have V limiting magnitudes fainter than 20 mag. For each source two panels are given: The right (with the ROSAT name on the top) is the V image with the dot size scaled according to the brightness with the scaling of V=15,17,19,21 shown in the upper left. In the right panel (with the limiting V magnitude on top) the size is scaled by the color (B-V=0.0,0.4,0.8 from left to right in the upper left scaling).

## 5 Conclusions

Our selection criterion according to the X-ray hardness ratio is very restrictive. An identical selection criterion has been applied to all sources in the *ROSAT* all-sky survey, and only about 30% of the known supersoft sources in the LMC and SMC were found (Greiner 1996). For instance, even a M31 source identical to CAL 83 would probably been missed with the present selection criterion. This is due to the fact that at lower intensities the error in the hardness ratio increases readily above 0.1–0.15 and removes the sources from the sample. Also, the selection criterion is more sensitive to low-temperature objects, i.e. sources with emission above 0.4 keV are already excluded. Therefore, a more relaxed hardness ratio selection will certainly increase the number of further sources which could qualify as supersoft sources.

The optically brightest supersoft X-ray source in the Large Magellanic Cloud has V=16.2 mag. A similar source in M31 is expected to have V=21.7 mag, still within the range of the *BVRI* survey. Those objects which have optical counterparts brighter than this might be expected not to be burning WDs in close binaries, but possibly Galactic single WDs or CVs or bright blue objects in M31. Also, a supersoft AGN is not excluded a priori as counterpart for X-ray sources in the outskirts of M31. We therefore conclude that not all of the sources described here in more detail may qualify as luminous close binaries like CAL 83 after spectroscopic identification has been succeeded.

The location distribution of our supersoft source sample across M31, especially the lack of sources in the inner bulge, clearly suggests that they belong to the disk population. Therefore, the detection probability is seriously affected by the absorbing column between the source and the observer.

*Acknowledgement:* JG is supported by the Deutsche Agentur für Raumfahrtangelegenheiten (DARA) GmbH under contract FKZ 50 OR 9201. The *ROSAT* project is supported by the German Bundesministerium für Bildung, Forschung, Wissenschaft und Technologie (BMBW/DARA) and the Max-Planck-Society.